\documentclass[12pt]{article}
\usepackage{graphicx}
\usepackage{amsmath}
\usepackage{floatflt}

\newcommand\pubnumber{CIPANP2015-Huber}
\newcommand\pubdate{\today}

\def\regina{Department of Physics\\
University of Regina, Regina, SK  S4S-0A2  CANADA}
\def\dal{Department of Physics and Atmospheric Science\\
Dalhousie University, Halifax, NS   B3H-4R2  CANADA}
\def\support{\footnote{Work supported by the Natural Sciences and Engineering
    Research Council of Canada (NSERC); FRN: 105851-2011 and SAPPJ-2015-00023.}}

\def\Title#1{\begin{center} {\Large #1 } \end{center}}
\def\Author#1{\begin{center}{ \sc #1} \end{center}}
\def\Address#1{\begin{center}{ \it #1} \end{center}}

\newcommand\pubblock{\rightline{\begin{tabular}{l} \pubnumber\\
         \pubdate  \end{tabular}}}
\newenvironment{Abstract}{\begin{quotation}  }{\end{quotation}}
\newenvironment{Presented}{\begin{quotation} \begin{center} 
             PRESENTED AT\end{center}\bigskip 
      \begin{center}\begin{large}}{\end{large}\end{center} \end{quotation}}
\def\Acknowledgements{\bigskip  \bigskip \begin{center} \begin{large}
             \bf ACKNOWLEDGEMENTS \end{large}\end{center}}

\def\beq{\begin{equation}}
\def\eeq#1{\label{#1}\end{equation}}
\def\eeqn{\end{equation}}

\def\beqa{\begin{eqnarray}}
\def\eeqa#1{\label{#1}\end{eqnarray}}
\def\eeqan{\end{eqnarray}}

\let\bar=\overbar

\def\Dslash{\not{\hbox{\kern-4pt $D$}}}
\def\dslash{\not{\hbox{\kern-2pt $\del$}}}

\def\msb{{\bar{\ssstyle M \kern -1pt S}}}

\begin{document}
\begin{titlepage}
\pubblock

\vfill
\Title{The Nucleon Polarizability Program at MAMI-A2}
\vfill
\Author{Garth Huber\support}
\Address{\regina}
\Author{Cristina Collicott}
\Address{\dal}
\Author{On behalf of the A2 Collaboration}
\vfill
\begin{Abstract}
Low energy Compton scattering allows the investigation of one of the
fundamental properties of the nucleon -- how its internal structure deforms
under an applied electromagnetic field.  We review recent developments in the
investigation of proton polarizabilities, and our plans for future measurement
at the Mainz Microtron (MAMI).
\end{Abstract}
\vfill
\begin{Presented}
Conference on the Intersections of\\
Particle and Nuclear Physics (CIPANP)\\
Vail, CO  USA  May 19--24, 2015
\end{Presented}
\vfill
\end{titlepage}
\def\thefootnote{\fnsymbol{footnote}}
\setcounter{footnote}{0}

\section{Introduction to Polarizabilities}

Nucleon polarizabilities are fundamental structure properties which are
sensitive to the internal quark dynamics of the nucleon.  They can be accessed
by measuring the differential cross section and singly and doubly polarized
asymmetries in Real Compton Scattering.  In this case, the low energy outgoing
Compton photon plays the role of the applied electromagnetic dipole field.  
In addition to their obvious interest as nucleon structure observables, nucleon
polarizabilities limit the precision that can be accessed in many other areas
of physics.  For example, in astrophysics, the polarizabilities influence the
properties of neutron stars.  In atomic physics, the polarizabilities yield an
appreciable correction to Lamb shift and hyperfine structure.  In fact, the
uncertainty in the proton's scalar polarizability is the largest uncertainty in
the proton radius extraction from the atomic hydrogen excitation spectrum
\cite{khrip98}.

The proton's electric and magnetic polarizabilities appear in the second order
term in the Compton scattering Hamiltonian
\begin{equation}
H^{(2)}_{eff}=\frac{1}{2}\alpha_{E1}\vec{E}^2+\frac{1}{2}\beta_{M1}\vec{H}^2,
\end{equation}
where $\alpha_{E1}=(11.2\pm 0.4)\times 10^{-4}$~fm$^{3}$ and
$\beta_{M1}=(2.5\pm 0.4)\times 10^{-4}$~fm$^{3}$ \cite{PDG2014}.
Despite their all-pervading nature, and great theoretical interest, there is
still a large uncertainty in the nucleon scalar polarizabilities.  $\alpha_{E1}$
is well constrained by experimental data, but $\beta_{M1}$ is less certain, and
the neutron data are particularly uncertain.

The spin polarizabilities appear in the third order term
of the effective interaction Hamiltonian
\begin{equation}
H^{(3)}_{eff}=\frac{1}{2}\bigl[\gamma_{E1E1}\vec{\sigma}\cdot\vec{E}\times\dot{\vec{E}}
+\gamma_{M1M1}\vec{\sigma}\cdot\vec{H}\times\dot{\vec{H}}
+2\gamma_{E1M2}H_{ij}\sigma_iE_j-2\gamma_{M1E2}E_{ij}\sigma_iH_j\bigr],
\end{equation}
involving one field derivative with respect to either time or space
$\dot{\vec{E}}=\partial_t\vec{E}$, $E_{ij}=1/2(\nabla_iE_j+\nabla_jE_i)$.
For example, $\gamma_{M1E2}$ represents the contribution where the proton is
excited by electric quadrupole $(E2)$ radiation and decays by magnetic dipole
$(M1)$ radiation.  The spin polarizabilities describe the ``stiffness'' of the
proton's spin against electromagnetic-induced deformations relative to the spin
axis, defining the frequency of the proton's spin precession induced by
variable electromagnetic fields.  Each spin polarizability is
dominated by a pion-pole contribution, the dispersive (interesting)
contribution is expected to be relatively small.

The proton's spin polarizabilities have never been measured, although the
forward and backward linear combinations $\gamma_0$ and $\gamma_{\pi}$ have
been.  The forward spin polarizability
$\gamma_0=-\gamma_{E1E1}-\gamma_{M1M1}-\gamma_{E1M2}-\gamma_{M1E2}$ can be
obtained from the polarized Compton cross section difference
$\sigma_{1/2}-\sigma_{3/2}$.  It is known to about 10\%, $\gamma_0=-(1.00\pm
0.08\pm 0.10)\times 10^{-4}$~fm$^4$ \cite{gamma0} and the pole contribution is
believed to cancel.  The backward spin polarizability
$\gamma_{\pi}=\gamma_{E1E1}+\gamma_{M1M1}-\gamma_{E1M2}+\gamma_{M1E2}$ is
obtained from unpolarized backward angle Compton scattering, and a recent
analysis gives the value as $\gamma_{\pi}=-(38.7\pm 1.8)\times 10^{-4}$~fm$^4$
\cite{gammapi}.  The pion contribution has been calculated as -46.7, so the
dispersive part $8.0\pm 1.8$ is known only to about 25\%.

\section{Measurements by A2 Collaboration at MAMI}

The polarizability program at MAMI is very active, with developments in
equipment, experiment and theory.  These are intended to produce a
comprehensive, consistent and precise data set, necessary for a reliable
extraction of the scalar polarizabilities of the proton and the neutron, and
the world's first independent extraction of the proton spin polarizabilities.
The A2 Collaboration utilizes a high-flux tagged bremsstrahlung photon beam,
liquid hydrogen or dynamically polarized butanol frozen spin target, and the
large acceptance Crystal Ball and TAPS detectors for these measurements.  For
more information, see Ref.~\cite{neiser}.

Until now, experiments relying on Disperson Relation (DR) analysis have only
been able to extract the sum and difference of $\alpha_{E1}$, $\beta_{M1}$,
resulting in correlated errors for the two quantities, and since the magnetic
polarizability is much smaller than the electric, the relative error on
$\beta_{M1}$ is large.  The desire for more precise proton $\beta_{M1}$ has
motivated a next generation experiment at MAMI-A2.  If combinations of cross
sections with {\em linearly polarized photon beam} are used, the leading-order
contributions from $\alpha_{E1}$, $\beta_{M1}$ are \cite{maximon}
\begin{equation}
\begin{aligned}
\frac{d\sigma^{\parallel}}{d\Omega} & =\frac{d\sigma^{\parallel}_{Powell}}{d\Omega}-
\frac{e^2}{2\pi m_p}\biggl(\frac{\nu^{\prime}}{\nu}\biggr)^2\nu\nu^{\prime}
(\alpha_{E1}\cos^2\theta+\beta_{M1}\cos\theta)+O(\nu^3)
\\
\frac{d\sigma^{\perp}}{d\Omega} & =\frac{d\sigma^{\perp}_{Powell}}{d\Omega}-
\frac{e^2}{2\pi m_p}\biggl(\frac{\nu^{\prime}}{\nu}\biggr)^2\nu\nu^{\prime}
(\alpha_{E1}+\beta_{M1}\cos\theta)+O(\nu^3).
\end{aligned}
\end{equation}
Thus, the contributions of the two scalar polarizabilities can in principle be
disentangled from measurements of the Compton angular distribution with
linearly polaried photon beam.  More recently, Krupina and Pascalutsa
\cite{kru2013} have shown that for energies below the $\Delta$ resonance, it is
better to use the polarized beam asymmetry $\Sigma_3$ to extract the
poorly-known $\beta_{M1}$
\begin{equation}
\Sigma_3\equiv\frac{d\sigma^{\perp}-d\sigma^{\parallel}}{d\sigma^{\perp}+d\sigma^{\parallel}}\\
=\Sigma_3^{Born}-f_3(\theta)\beta_{M1}\nu^2+O(\nu^4).
\label{eqn:sig3}
\end{equation}


Following this motivation, we recently measured, for the first time, the beam
asymmetry $\Sigma_3$ below pion production threshold ($E_{\gamma}=80-140$~MeV).
More than 70.000 $\gamma p\rightarrow\gamma p$ events for each of the two
polarization settings were obtained, with an overall background contamination
below $5\%$. The preliminary data are in a good agreement with Chiral
Perturbation Theory ($\chi$PT) \cite{kru2013} and Heavy Baryon Chiral
Perturbation Theory (HB$\chi$PT) \cite{mcg2013} calculations, however, a
noticeable deviation from the Born term, independent of proton polarizabilities
(see Eqn.~\ref{eqn:sig3}) was observed.  The data will be presented, along with
the corresponding estimates for the proton scalar polarizabilities, in an
upcoming publication \cite{sokhoyan2015}. So far, only 1/3 of the approved data
were taken.  The remaining 2/3 of the data are expected to be acquired in 2016,
after an upgrade of the Glasgow-Mainz tagger to allow four times higher rate
compared to the already performed measurement.

Since the spin polarizabilities appear in the
effective interaction Hamiltonian at third order in photon energy, they are a 
small effect at lower energies.  It is in the $\Delta$ resonance region
($E_{\gamma}=200-300$ MeV) where their effect becomes significant.  In this
energy region, it is possible to accurately measure polarization asymmetries
using a variety of polarized beam and target combinations.  The various
asymmetries respond differently to the individual spin polarizabilities at
different $E_{\gamma}$ and $\theta$, so it is by measuring at least three
different asymmetries at different $E_{\gamma}$, $\theta$ that their
contributions can be isolated.  The plan of the A2 Collaboration is to conduct
a global analysis, including constraints from all available prior data
(e.g. $\alpha_{E1}$, $\beta_{M1}$, $\gamma_0$, $\gamma_{\pi}$) to independently
extract all four spin polarizabilities with small statistical, systematic and
model-dependent errors.

The $\Sigma_3$ single-spin asymmetry with linearly polarized beam has been
defined above.  The two double-spin asymmetires of interest with circularly 
polarized beam are defined as
\begin{equation}
\Sigma_{2z}=\frac{\bigl(\frac{d\sigma}{d\Omega}\bigr)_{\uparrow\uparrow}-\bigl(\frac{d\sigma}{d\Omega}\bigr)_{\uparrow\downarrow}}
{\bigl(\frac{d\sigma}{d\Omega}\bigr)_{\uparrow\uparrow}+\bigl(\frac{d\sigma}{d\Omega}\bigr)_{\uparrow\downarrow}}
\ \ \ \ \ 
\Sigma_{2x}=\frac{\bigl(\frac{d\sigma}{d\Omega}\bigr)_{\uparrow\rightarrow}-\bigl(\frac{d\sigma}{d\Omega}\bigr)_{\uparrow\leftarrow}}
{\bigl(\frac{d\sigma}{d\Omega}\bigr)_{\uparrow\rightarrow}+\bigl(\frac{d\sigma}{d\Omega}\bigr)_{\uparrow\leftarrow}},
\end{equation}
where the second $\uparrow\downarrow$ indicate transversely polarized nucleon
target orientation, and the $^\rightarrow_\leftarrow$ indicate longitudinally
polarized target.


These measurements are quite challenging.  The Compton scattering cross section
is small, only about 1\% of the dominant $\pi^0$ photoproduction process at
these energies, and under certain conditions $\pi^0$ photoproduction can mimic
the Compton scattering signature if one of the photons escapes the detector or
if the electromagnetic showers from the two photons overlap (due to finite
angular resolution).  In addition, coherent and incoherent reactions from C, O,
He in the polarized butanol target need to be identified and subtracted.  In
the $\Delta$-region, the use of the recoil proton track can sometimes assist in
the suppression of non-Compton backgrounds, but the energy losses in the target
elements and inner detector elements are considerable and greatly restrict the
kinematic region where the recoil proton track provides reliable information.

\begin{figure}[htb]
\centering
\includegraphics[width=2.9in]{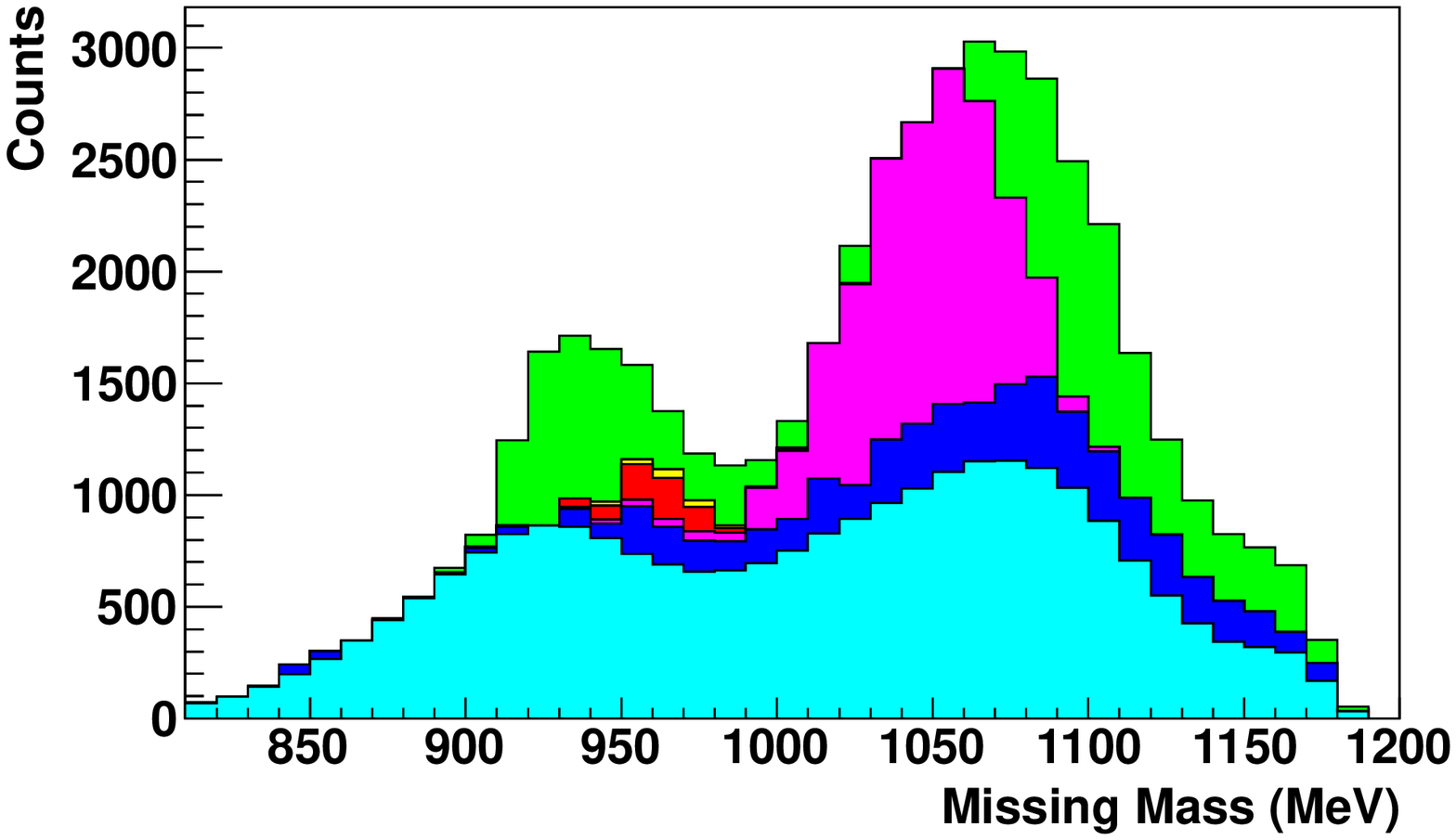}
\includegraphics[width=2.9in]{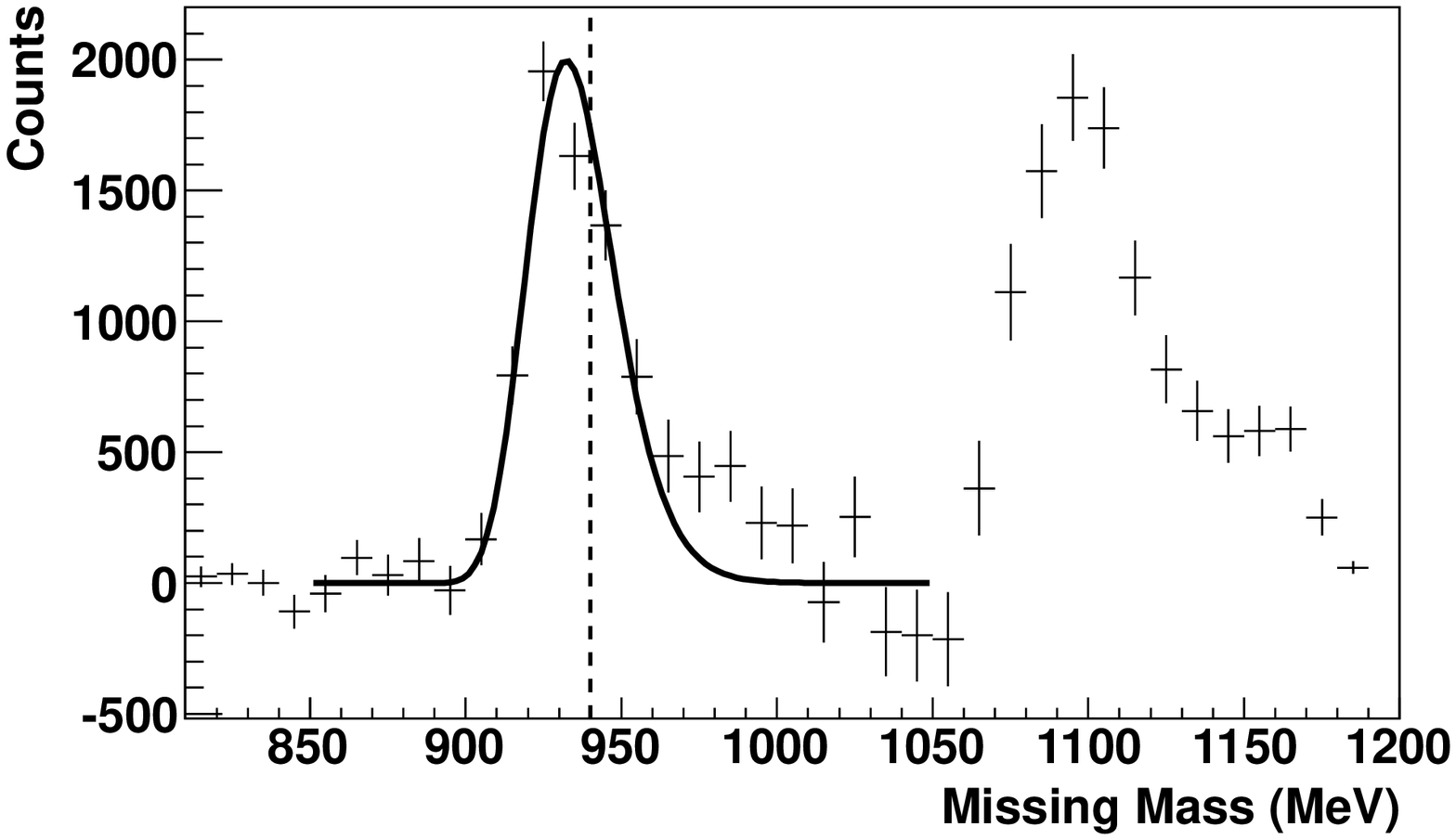}
\caption{{\bf Left:} Data Missing Mass distribution from $\Sigma_{2x}$ analysis
  for $E_{\gamma}=273-303$ MeV, $\theta_{LAB}^{\gamma^{\prime}}=100-120^o$
  \cite{martel2015} 
  (green) with various background contributions indicated as
  follows: accidental coincidences (cyan), carbon/cryostat contributions
  (blue), reconstructed $\pi^0$ background where one decay $\gamma$ escapes the
  setup via the TAPS downstream hole (red), or the Crystal Ball upstream hole
  (magenta).
{\bf Right:} Fully-subtracted Missing Mass spectrum with simulated Compton peak
overlaid.  A conservative MM$<$940 MeV cut is then applied to exclude $\pi^0$
production.
\label{fig:MM_martel}}
\end{figure}

\begin{figure}[htb]
\centering
\includegraphics[width=2.9in]{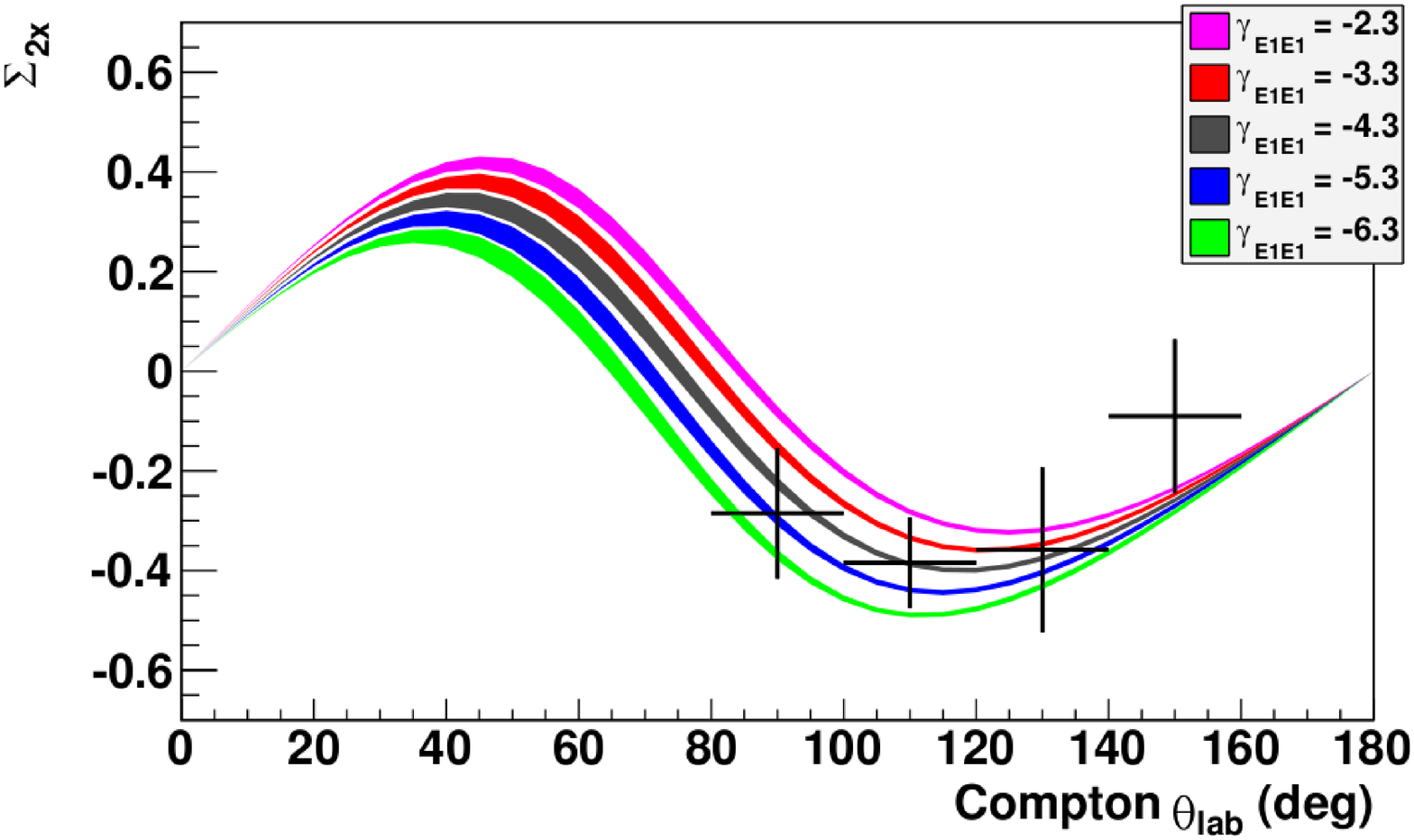}
\includegraphics[width=2.9in]{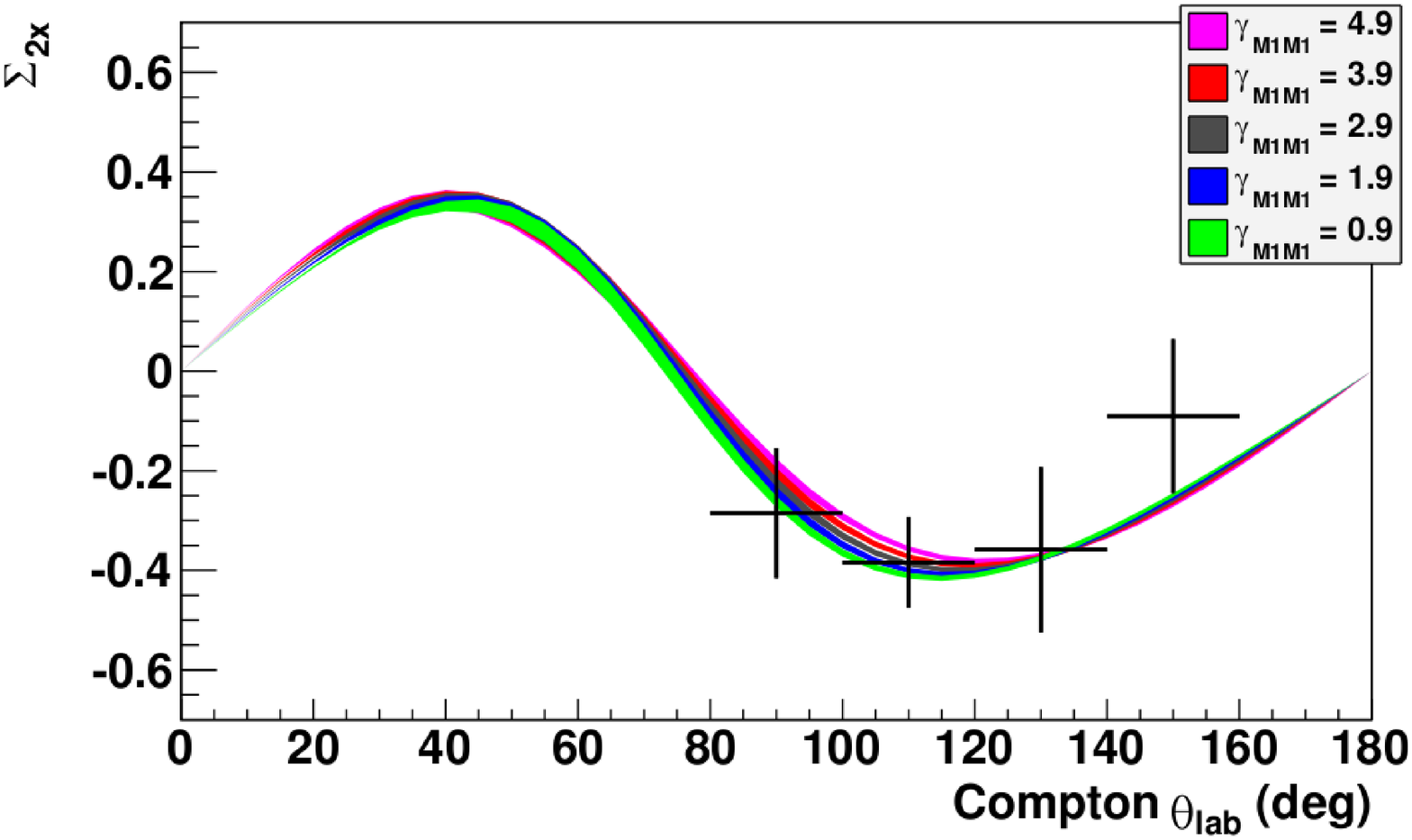}
\caption{$\Sigma_{2x}$ results for $E_{\gamma}=273-303$ MeV versus
  $\theta_{LAB}^{\gamma^{\prime}}$ \cite{martel2015}.  Overlaid are DR 
  calculations of Pasquini et al. \cite{pasquini} making use constraints on
  $\alpha_{E1}+\beta_{M1}$, $\alpha_{E1}-\beta_{M1}$, $\gamma_0$,
  $\gamma_{\pi}$ (allowed to vary within experimental errors).
 {\bf Left:} $\gamma_{M1M1}$ is fixed in the calculation and $\gamma_{E1E1}$ is
 varied, as indicated.  {\bf Right:} $\gamma_{E1E1}$ is fixed and
 $\gamma_{M1M1}$ is varied.
\label{fig:Sig2x}}
\end{figure}

Fig.~\ref{fig:MM_martel} provides an example of these background contributions
from our recent $\Sigma_{2x}$ analysis \cite{martel2015}.  After the indicated
cuts are applied, a clean data sample is obtained.  Our first measurement
of a double-spin Compton scattering asymmetry on the nucleon is shown in
Fig.~\ref{fig:Sig2x}.  As indicated, the results are clearly sensitive to the
value of $\gamma_{E1E1}$ in the calculation, but not very sensitive to the
value of $\gamma_{M1M1}$.  Additional data to further reduce the $\Sigma_{2x}$
error bars are planned to be acquired in 2017.

\begin{figure}[htb]
\centering
\includegraphics[width=2.9in]{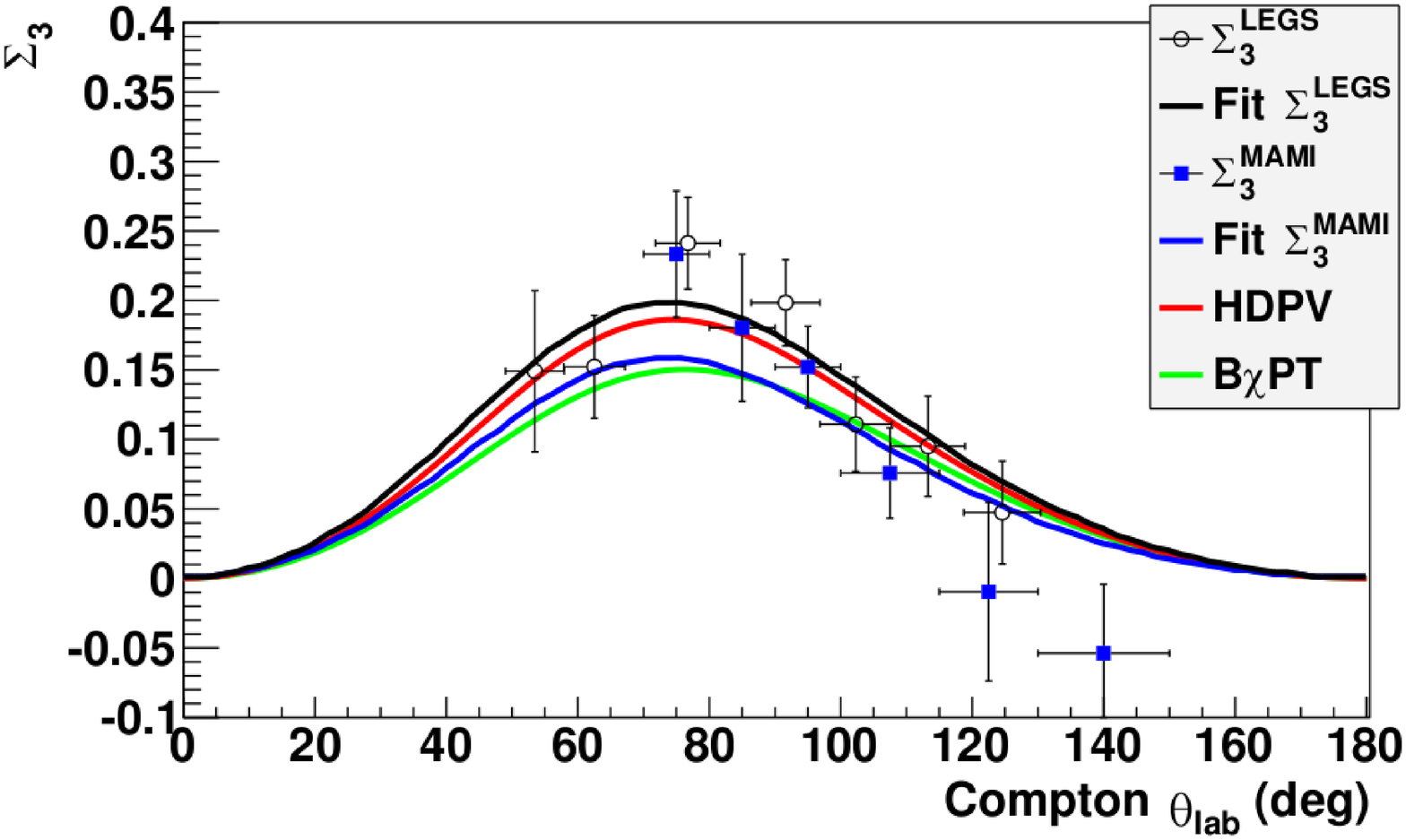}
\includegraphics[width=2.9in]{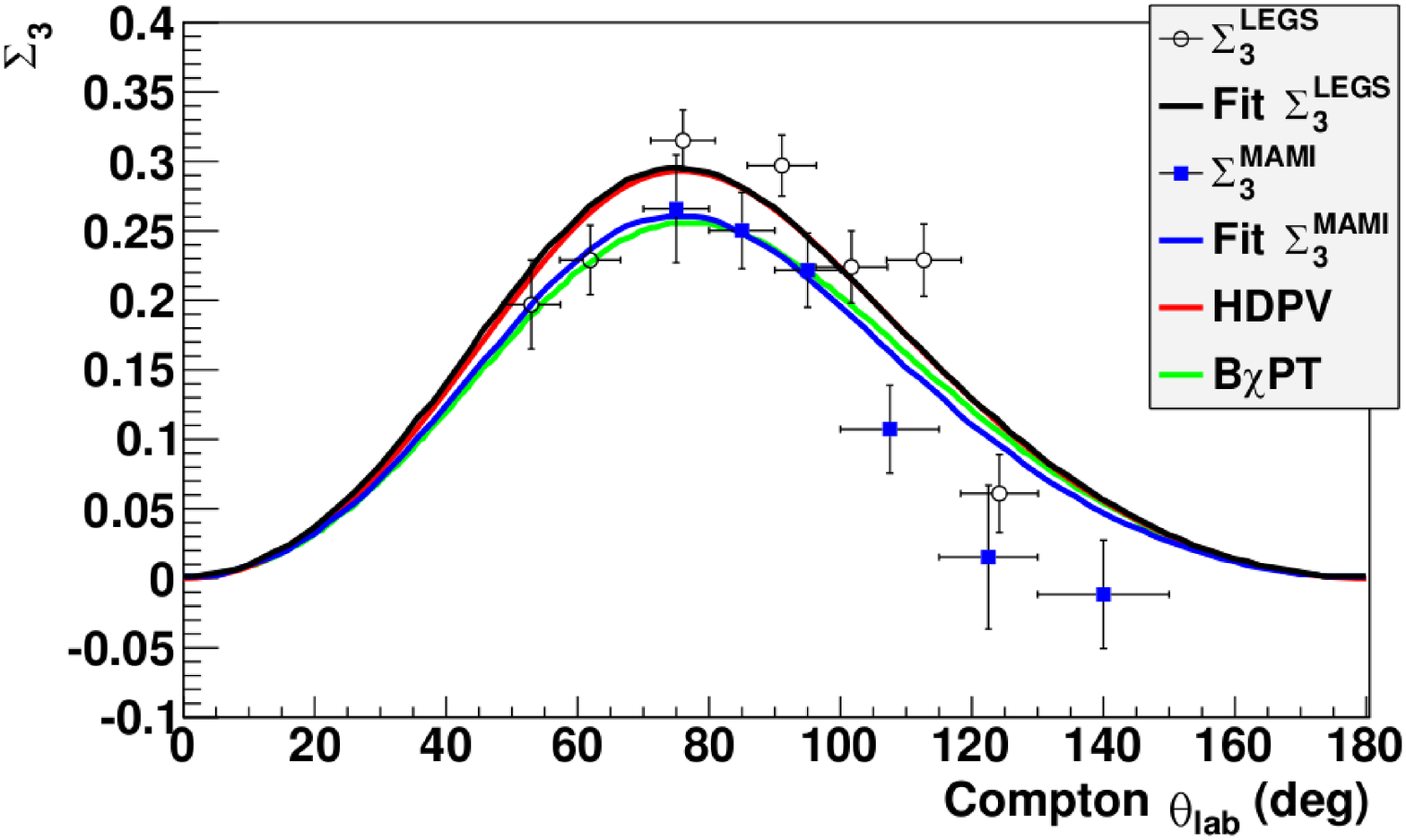}
\caption{Preliminary $\Sigma_{3}$ results for {\bf Left:}
 $E_{\gamma}=267-287$ MeV, {\bf Right:} $E_{\gamma}=287-307$ MeV,
  versus $\theta_{LAB}^{\gamma^{\prime}}$ \cite{collicott}, in comparison with
  previously published data from LEGS \cite{legs}.  Overlaid are DR
  calculations of Refs. \cite{pasquini, bchipt} using their preferred
  polarizabilities.  Only statistical uncertainties are shown.
\label{fig:Sig3}}
\end{figure}

In Fig.~\ref{fig:Sig3} are our preliminary results for $\Sigma_3$ in
the resonance region \cite{collicott}.  Similar to Fig.~\ref{fig:MM_martel},
restrictive cuts to eliminate $\pi^0$ and other backgrounds are applied,
resulting in a clean data sample.  The resulting missing mass distribution
agrees well with our Compton scattering simulations, and the extracted
$\Sigma_3$ for $\pi^0$ photoproduction also agrees well with previously
published data.  Our $\Sigma_3$ results are also compared to those from the
LEGS Collaboration \cite{legs} in Fig.~\ref{fig:Sig3}.  Even though the
statistical uncertainties from both measurements are rather large, a shift in
the asymmetries can be observed, particularly near $90^o$ in the higher energy
region.  The MAMI results suggest that the $\Sigma_3$ asymmetry may fall off
more rapidly than predicted at backward angles.

A recent analysis by Martel, et al. \cite{martel2015} determined the proton's
spin polarizabilities for the first time, combining $\Sigma_3$ results from the
LEGS Collaboration \cite{legs} and $\Sigma_{2x}$ results from MAMI.  The
analysis used a fixed-$t$ DR code, provided by B. Pasquini \cite{pasquini}, to
fit the asymmetry data.  The fitting routine varies $\alpha+\beta$,
$\alpha-\beta$, $\gamma_0$, $\gamma_{\pi}$ and $\gamma_{E1E1}$,
$\gamma_{M1M1}$, to fit the asymmetry data.  The first four are allowed to vary
only within their known experimental uncertainties.  These results are shown in
the third column of Table \ref{tab:spin_pol}.

\begin{table}[t]
\begin{center}
\begin{tabular}{l|cc|c|c}  
  & HDPV \cite{pasquini} & B$\chi$PT \cite{bchipt} & $\Sigma_{2x}$ and
  $\Sigma_3^{LEGS}$ \cite{martel2015} & $\Sigma_{2x}$ and
  $\Sigma_3^{MAMI}$ \cite{collicott} \\ \hline
 $\gamma_{E1E1}$ & -4.3 & -3.3 & -3.5$\pm$1.2   & -5.0$\pm$1.5  \\ \hline
 $\gamma_{M1M1}$ &  2.9 &  3.0 &  3.16$\pm$0.85 & 3.13$\pm$0.88 \\ \hline
 $\gamma_{E1M2}$ & -0.0 &  0.2 & -0.7$\pm$1.2   &  1.7$\pm$1.7  \\ \hline
 $\gamma_{M1E2}$ &  2.2 &  1.1 &  2.0$\pm$0.3   &  1.3$\pm$0.4  \\ \hline
 $\gamma_0$     & -0.8 & -1.0 & -1.03$\pm$0.18 & -1.00$\pm$0.18 \\ \hline
 $\gamma_{\pi}$  &  9.4 &  7.2 &  9.3$\pm$1.6  &  7.8$\pm$1.8 \\ \hline
 $\alpha+\beta$ &      &      &  14.0$\pm$0.4 & 13.8$\pm$0.4 \\ \hline
 $\alpha-\beta$ &      &      &   7.4$\pm$0.9 &  6.6$\pm$1.7 \\ \hline
 $\chi^2/d.f.$  &      &      &          1.05 &          1.25\\ \hline

\end{tabular}
\caption{Disperson Relation fits to $\Sigma_{2x}$ along with either
  $\Sigma_3^{LEGS}$ or $\Sigma_3^{MAMI}$.  Also shown are the preferred values
  from two models.  Spin polarizabilities are in units of $10^{-4}$ fm$^4$, and
  scalar polarizabilities are in units of $10^{-4}$ fm$^3$.
\label{tab:spin_pol}}
\end{center}
\end{table}

The LEGS cross section data have some significant discrepancies when compared
to other data sets \cite{camen}.  Because it is possible that a discrepancy
exists only in the cross sections and not the asymmetries, it is worthwhile to
check the sensitivity of the extracted spin polarizability results to these
data.  In this case, the same fitting algorithm applied in the Martel, et
al. \cite{martel2015} analysis was applied also to the new MAMI $\Sigma_3$
results, resulting in the values shown in the fourth column of
Table \ref{tab:spin_pol}.  This is the first extraction of all four spin
polarizabilities using only MAMI data.

It is important to note that the LEGS $\Sigma_3$ data set covers a wide angular
and energy range and consists of 58 data points, while the MAMI $\Sigma_3$ data
are only 12 data points.  A comparison of the two sets of $\gamma_i$ values show
that the errors on the individual spin polarizabilities increase slightly
when using only MAMI data.  However, considering the reduced data set in
comparison to LEGS, this is to be expected.  While the value of $\gamma_{M1M1}$
remains nearly unchanged, a significant shift is seen in the other three spin
polarizabilities.  Note in particular the significant shift in $\gamma_{\pi}$.
It has been noted previously \cite{camen} that the LEGS data show a large
discrepancy from all other data sets when used to extract $\gamma_{\pi}$.  This
discrepancy is further confirmed here.  Because $\gamma_{E1M2}$ and
$\gamma_{M1E2}$ are determined through their linear relation to $\gamma_0$,
$\gamma_{\pi}$, a large shift in $\gamma_{\pi}$ helps to explain the differing
spin polarizabilities.

In summary, the polarizabilities program at MAMI is very active, providing new
data for testing QCD via Chiral Perturbation Theory and Dispersion Relations in
the non-perturbative regime.  For the scalar polarizabilities, we have
demonstrated a new technique to extract $\beta_{M1}$ from the $\Sigma_3$
asymmetry without correlated errors to the larger $\alpha_{E1}$.  Additional
data to complete this measurement are planned for 2016.  Regarding the spin
polarizabilities, we have embarked on a three part program to extract all four
spin polarizabilities independently with small statistical, systematic and
model-dependent errors.  We have measured the $\Sigma_{2x}$ double-spin
asymmetry for the first time, and plan to acquire additional statistics in the
next 2 years.  We have taken new $\Sigma_3$ data to supplement existing data,
and used both asymmetries to extract the four spin polarizabilities for the
first time.  Our planned $\Sigma_{2z}$ data will further constrain the spin
polarizabilities.  The first set of these data were acquired in 2014, and we
have just concluded a very successful second run in the summer of 2015.
Finally, the A2 Collaboration plans to eventually extend the polarizability
measurements to the neutron, using a high-pressure active $^3$He gas
scintillator target \cite{annand}.

\Acknowledgements

We wish to thank the MAMI accelerator group and staff for their outstanding
support, and the CIPANP conference organizers for the talk invitation.


\begin{thebibliography}{99}


\bibitem{khrip98} I.B. Khriplovich, R.A. Sen'kov, Phys. Lett. A {\bf 279}, 474 (1998).
\bibitem{PDG2014} K.A. Olive, et al. (Particle Data Group), Chin. Phys. C {\bf
  38}, 090001 (2014).
\bibitem{gamma0} J. Ahrens, et al. (GDH and A2 Collaborations),
  Phys. Rev. Lett. {\bf 87}, 022003 (2001).
\bibitem{gammapi} M. Schumacher, Prog. Part. Nucl. Phys. {\bf 55}, 567 (2005).
\bibitem{neiser} A. Neiser (A2 Collaboration), J. Phys. Conf. Ser. {\bf 587},
  012041 (2015).
\bibitem{maximon} L.C. Maximon, Phys. Rev. C {\bf 39}, 347 (1989)
\bibitem{kru2013} N. Krupina, V. Pascalutsa, Phys. Rev. Lett. {\bf 110}, 262011
  (2013).
\bibitem{mcg2013} J. McGovern, D. Phillips, H. Griesshammer, Eur. Phys. J. A
  {\bf 49}, 12 (2013).
\bibitem{sokhoyan2015} V. Sokhoyan, et al. (A2 Collaboration), to appear.
\bibitem{martel2015} P. Martel, et al. (A2 Collaboration),
  Phys. Rev. Lett. {\bf 114}, 112501 (2015), arXiv:1408.1576.
\bibitem{pasquini} B. Pasquini, et al., Phys. Rev. C {\bf 76}, 014203 (2007).
\bibitem{collicott} C. Collicott, et al. (A2 Collaboration), to appear.\\
C. Collicott, Ph.D. thesis, Dalhousie University, 2015.\\
http://dalspace.library.dal.ca/handle/10222/56762
\bibitem{legs} G. Blanpied, et al. (LEGS Collaboration), Phys. Rev. C {\bf 64}
  025203 (2001).
\bibitem{bchipt} V. Lensky, J. McGovern, Phys. Rev. C {\bf 89} 032202 (2014).
\bibitem{camen} M. Camen, et al., Phys. Rev. C {\bf 65}, 032202(R) (2002).
\bibitem{annand} J. Annand, et al. (A2 Collaboration), MAMI Proposal A2-01-2013.

\end{thebibliography}
\end{document}